# Quantum Vacuum Charge and the New HyperCP Particle X

*© 2005 C.F. Diether III[1*], A.E. Inopin[2*]*

[1]*Phaedo Engineering, www. vacuum-physics, com, Arleta, CA, USA*
[2]*Kharkov National University, Department of Experimental Nuclear Physics,
Svobody Sq. 4, 61077, Kharkov, Ukraine*

We present a unique geometrical model based on our findings of a quantum vacuum charge scenario to predict the recent experimental find of the new HyperCP 214 MeV particle state X. This model, which we call the Spin Matrix, also predicts many more states; some of which represent known particles and some states that aren't represented directly by known particles or haven't been experimentally found yet. We demonstrate a parameter-free description of the lowest energy particles in nature.

## 1. Introduction

During this year, 2005, the HyperCP collaboration at Fermilab announced a new rare particle at approximately 214 MeV that was found in their data [1]. This find allowed us to modify our Spin Matrix geometry model of the organized part of the quantum vacuum that we presented in a previous paper to better fit existing data [2]. Starting with the rest mass energy of a positron we predict the rest mass energies of the up and down quarks, the $\pi^0$ meson and now this new X particle at 214 MeV using our geometrical model modified by a fitting function. We also predict other energy states that don't directly relate to elementary particles of the Standard Model but they possibly do relate to higher resonances of composites.

In this paper we will discuss more details about our Spin Matrix geometrical model and how its basic linear geometrical configuration must be modified for higher energy non-linearities. We use the term Dual Spacetime to imagine this concept and Spin Matrix for the organized part of the quantum vacuum that is mostly hidden from us. Our model is much like a modified Dirac Sea that involves two 3-branes connected by a common 2D surface boundary. This type of concept has become popular in recent years with the advent of the Randall-Sundrum papers [3].

Another very important feature of our Spin Matrix geometry is chiral symmetry and how it is broken. And maybe a clue to how chiral symmetry is restored at higher energies. It will be easy to see that there is a mix of energy states that probably have chiral symmetry and some that don't. But first we need to say that our energy states represented by the circular areas are composites of fermion / anti-fermion pairs resulting from our perspective that they are much like Goldstone bosons with zero overall spin and zero mass. However, we must also take heed to the *Anyon* hypothesis and that some of these energy states might be neither fermionic nor bosonic in content. Which leads to fractional statistics [4]. For the most part of this letter though, we will assume that the content of the energy states to be fermionic / anti-fermionic pairs which ought to behave similar to Cooper pairs from BCS theory [5].

And of course, this Spin Matrix geometry can explain why there is a mass gap in the first place since it only allows certain energy states to exist due to its interactional equilibrium. We can easily see from the geometry that e+e- pairs are the first to come out of the Spin Matrix into our spacetime and then pions with respect to first generation particles. We must admit that we don't seem to see the presence of 2$^{nd}$ or 3$^{rd}$ generation fermions in this scenario yet. But we expect that a less trivial geometric scheme must produce them.

---

[*] E-mail addresses: fredifizzx@hotmail.com (Diether), inopin@yahoo.com (Inopin).

## II. The Spin Matrix Geometry

In the previous paper, we also presented the concept of quantum vacuum charge (QVC) = $\pm\sqrt{\hbar c}$ and we showed that it works very well for the bound charge of free space photons [2]. This is what led us to the concept of the Spin Matrix geometry. For QVC to be a reality, then there must be some kind of quantum mechanism to support it. One naturally thinks of virtual fermionic pairs surrounding real charged elementary particles and QVC is an extension of that screening. When we say "fermionic pair" we mean with a fermion and anti-fermion of the same particle flavor.

Our Spin Matrix geometrical model of the quantum vacuum is fairly simple and is based on the idea that the quantum vacuum should have properties that dictate what we see as the real elementary particle spectrum. In other words, an organization amidst the quantum chaos. It is strictly two-dimensional and can be seen in the figures that follow. It is basically a Dirac-like Sea concept except that only certain energy states are allowed by the geometry instead of having a continuum of possible energy states. Now in thinking of a Dirac Sea concept, we have to imagine that the "Sea" would have to be composed of "less than virtual" (LTV) fermionic pairs. Of course this presents the usual problem of *where is this sea of LTV fermions*? We think it is mostly hidden from us on the other side of an event horizon. In other words, there are two 3-branes connected by a common 2D surface or event horizon. Our 2D Spin Matrix is the view of looking from our brane into the brane on "the other side of the looking glass" on a very microscopic level. This concept of a sea of LTV fermionic pairs, immediately allows for virtual fermionic pairs to exist. Virtual fermionic pairs are simply the state of being between both "worlds". When we say "less than virtual" we mean the set of particles that are in the other brane that we can have no direct contact with because they are beyond the event horizon. This is similar to the concept that we can have no direct contact with particles that might be in a black hole.

The dual space-time scenario that we are thinking of would have to be similar to a Dirac Sea only modified so that it can work. And we must turn to cosmology for a more complete description to help with this. There is really good evidence that there was some kind of Big Bang event that started *our* Universe so it is easy to imagine that there is an event horizon that is associated with our Universe's "now". This event horizon must be physically moving at c right now. This gives us a necessary boundary between the two spacetimes. And better allows for a modified Dirac Sea concept to work.

Figure1 shows a two dimensional hexagonal configuration of the Spin Matrix. We can see that this produces areas that are "isolated" from each other in the space where the 3.3 MeV energy states form and this immediately brings to mind regions where the "weak" and "strong" interactions operate. And that higher energy states can form in the smaller spaces. The hexagon outlined in Fig.1 would be a "cell" of the Spin Matrix. We call this cell a *zerton* and it should have the property of $\pm\sqrt{\hbar c}$ for its overall charge. It can also be seen that the electromagnetic interaction is not isolated and this can explain its "infinite" range. Whereas the smaller areas, being isolated from each other, have interactions that are limited in range. It looks like there is the possibility of *tunneling* between the isolated regions though. However, the mechanism of how all of this works in our spacetime from the Spin Matrix spacetime needs more explanation.

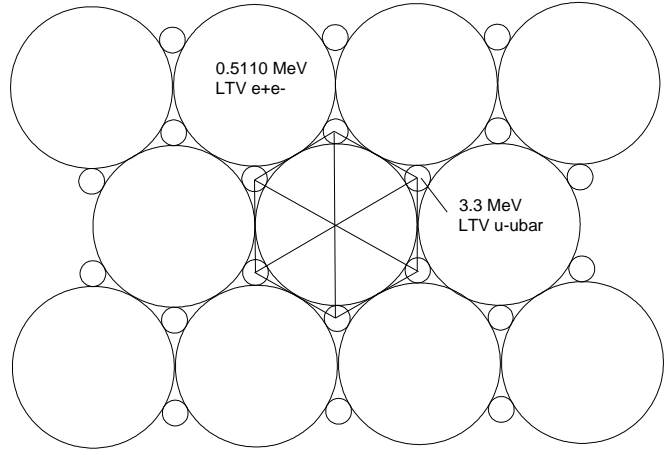

Fig. 1. Hexagonal configuration of the 2D Spin Matrix boundary. If we take the larger areas to be 0.511 MeV, then the smaller areas shown, geometrically come out to be 3.3 MeV.

All particles that we consider to be real in our spacetime have to have a *connection* to the Spin Matrix. This connection would be the cloud of virtual pairs that surround all fermions.

In Figure 2 we have zoomed in on one of the *isolated* regions. To calculate the energy of the energy states, we set the radius of the LTV $e^+e^-$ state equal to 1 and then just divide $m_e c^2$ by the radius of the smaller circles that is calculated from the geometrical relationships to obtain an energy value for them. Basically we are just making a relation of these states using the reduced mass formula,

$$m_{\text{red}} = \frac{m_1 m_2}{m_1 + m_2} = \frac{m}{2} \text{ when } m_1 = m_2, \tag{1}$$

similar to how a bound state like positronium would be considered so that we can have a value of what this state might have in relation to a value in our spacetime to a single fermion of the pair. From our macroscopic perspective, these states really look to have zero energy so we must use a microscopic perspective by just starting with what the rest mass energy of an electron or positron would be and scale the rest from that. In other words, this is what a real fermion might "see" with respect to energy values. The LTV $e^+e^-$ from our perspective is the annihilated state of positronium. An example of obtaining the energy for the state of the next smaller circle is that we can calculate the radius of the circle by geometry and we obtain $2\sqrt{3}/3 - 1 \cong 0.15470$. Then $m_e c^2/(2\sqrt{3}/3 - 1) \cong 3.303$ MeV for the result of the next smaller circle. And so on as we go to smaller and smaller *sized* energy states (actually higher energy). In theory it is possible to obtain exact geometrical results. However, because of the quantum nature of the Spin Matrix, we would expect that what we are presenting are average values anywise.

In Figure 3, we obtain energy states that are close to a $\pi^0$ meson and then the HyperCP 214 MeV X particle. The actual linear geometrical calculations for the $\pi^0$ meson area is 142.7 MeV and for the HyperCP X is 235.7 MeV. At this point we make the conjecture that these energy states do represent these particles and we apply a "fitting function" to adjust the values of the Spin Matrix geometry. We will discuss this function more later but basically we think it is necessary because of higher energy non-linearities. So all of the values we have labeled on the figures are the adjusted values. We think this "fitting" is due to the concept that the lower energy states are *softer* and higher energy states are *harder* so that in reality what we are showing as perfect circles really aren't that perfect in the Spin Matrix quantum mechanism. Of course we also have the problem to figure out of why the 134.7 MeV state is "picked out" to represent the $\pi^0$ meson. There must be some defining reason why the $u\bar{u}$ and $d\bar{d}$ quarks pick this *current path* when being excited into our spacetime from the Spin Matrix. We also note that we have an unexplain state of 14.93 MeV in Figure 2. We think this state is a new lightest Goldstone state and is yet to be discovered.

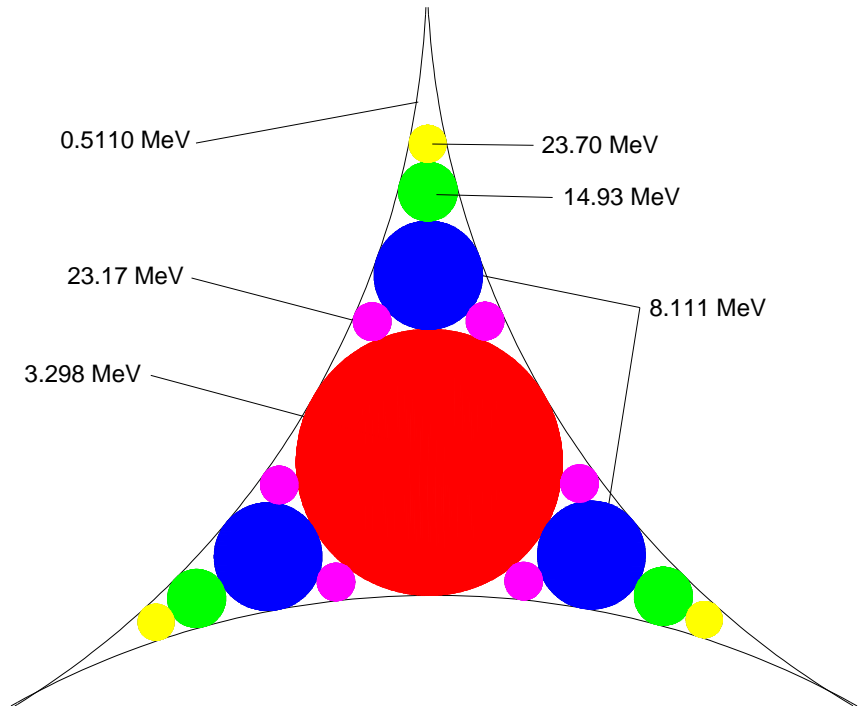

Figure 2. Zoom in of isolated region

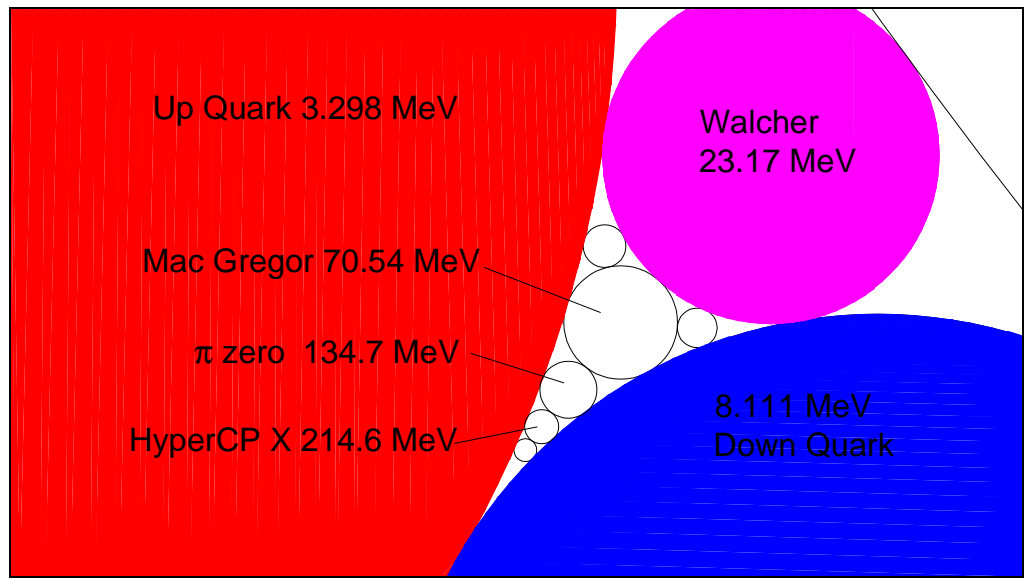

Figure 3. Further zoom in of small area between the LTV $u\bar{u}$ and $d\bar{d}$ quark areas.

## III. The HyperCP 214 MeV X Particle

Recently He, et al, has published results from HyperCP Collaboration [6]. The striking result is the discovery of a new particle X, with mass of about 214 MeV. The authors claim an X to be narrow, short-lived particle, and non-hadronic. They find that existing data on $K+ \rightarrow \pi^+\mu^+\mu^-$ rule out a scalar particle or a vector particle as possible explanations. The authors also explored all the existing constraints on pseudoscalar and axial vector particles, and conclude that these possibilities are still allowed. Another paper further constrains the HyperCP find to be a pseudoscalar, which we tend to agree with [7].

One of the possible interpretations for new X-particle is to be an sgoldstino, a mystical object, which arises in supersymmetrical schemes [1,6,7]. The sgoldstino has spin 0 and it can have mass, although this mass can be quite arbitrary. The sgoldstino is the bosonic superpartner for the goldstino. But there could be different scenarios for such physics, and we will suggest one of them. We have used our Spin matrix model to calculate masses of inner energy states originating from 2D structure in Figure 1 and found an energy state of the appropriate magnitude for the HyperCP 214 MeV particle lying right next to an energy state that we associate with pions as can be seen in Fig. 3. Our mass evaluations and predictions are essentially parameter-free.

## IV. X-particle Physics

Some time ago the group of authors – Akers, Palazzi, and Mac Gregor came up with the so-called "An Elementary Particle Constituent-Quark Model"[8]. They compared meson and baryon spectra and found a common mass-band structure, in which particles with a variety of quantum numbers are grouped together. These quantized mass values seemingly mandate a constituent-quark (CQ) viewpoint. By studying the meson and baryon yrast states, the authors deduced the masses and spin values of some of the constituent quarks. The authors' basic claim is that any baryon or meson mass could be simply expressed in terms of either X1 = 35 MeV or X2 = 70 MeV multiple quanta. One can see how nicely Mac Gregor's 70 MeV quanta fits in our geometrical solution (see Fig.3).

Another novel model was devised recently by Walcher [9]. This model describes the possible existence of narrow excited states in the nucleon sector below the πN threshold. A model based on the excitation of collective states of the quark condensate is proposed explaining these states as multiple production of a "genuine" Goldstone boson with a mass of ≈ 20 MeV. Walcher assumes that the π observed in nature is not the Goldstone boson that it is normally identified with, but a mixture of the "genuine" Goldstone boson $\pi_G$, i.e. the light π suggested above consisting of light current quarks, with a "constituent" $\pi_c$ made of constituent quarks with a mass of $m_{\pi c} = 2m_c = 700$ MeV. Despite many promises, this model suffers a few drawbacks in considering the dynamics of quarks and pions. We see how naturally Walcher's 22 MeV Goldstone arises in our Spin Matrix construction (see Figs 2, 3).

## IV.a "Chiral Asymptote"

We have obtained a universal fitting function from adjusting our linear Spin Matrix state values to fit the $\pi^0$ meson and the HyperCP 214 MeV energy, which works well for our purposes

$$f(x) = \frac{m_e c^2 x}{m_e c^2 + 0.000213 x}, \qquad (2)$$

where x equals the energy of the state from the Spin Matrix to be adjusted. We can see that eq.(2) is essentially a modification of the reduced mass eq. (1). We would like to note that our factor of 0.000213 is close to $4\alpha^2$ though we have not made any definitive connection yet. Let us see Figure 4 for the detailed behavior of our f(x). It is a nearly linear function at $0 < x < 100$ MeV, then it gets some non-linearity which increases with energy and starts to flatten out in the 200 MeV region. With energy increase, f(x) continues to behave this way further, until it is completely flattened out at about 2.4 GeV. But this is just a boundary of a region, which fits nicely into color screening or a partial confinement scenario. A few papers by Arbuzov [10], Goldman [11], Garcilazo [12], Jaffe [13], Stepanov [14] describe this rich phenomenon via different models, but our 2.4 GeV value just fits in the overall parameter's window. Garcilazo et al used a screened potential of the Bhaduri-type [12] and explained the so-called missing resonance problem in N-Δ spectra with screening boundary of just about 2.4 GeV.

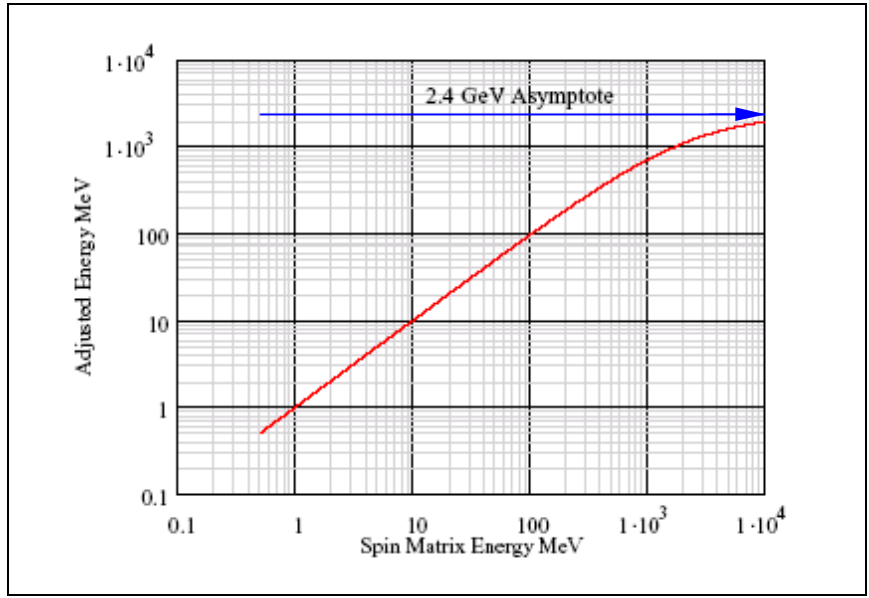

Figure 4. Fitting Function

Restoration of chiral symmetry can really be put on solid ground, which was done recently in ℏ-expansion GNJL approach by Kalashnikova et al [15]. The grand result shown is – chiral symmetry is restored high in the spectrum of heavy-light mesons, and indeed, parity doubling was demonstrated to occur above the scale of about 2.5 GeV, regardless of the explicit form of the confining potential. Kalashnikova [15] defines this scale of symmetry restoration $\Lambda_{rest}$, as a scale at which splitting within a chiral doublet is much smaller than the BCS scale $\Lambda_{BCS} = 250$ MeV. They also conclude that chiral symmetry is restored in the spectrum if the pionic wave function vanishes for given values of relative inter-quark momentum. But this is practically our asymptotic value of 2.4 GeV, which could be considered as a point of phase transition in our framework. The authors [15] identified the following three regimes 1) chiral regime, $p < \Lambda_{BCS}$, with chiral symmetry breaking playing a dominant role in the interaction; 3) the restoration regime, $\Lambda_{rest} < p$ (p-momentum), which realizes for highly excited states in spectrum, where parity doublets, and even chiral quartets appeared; 2) the intermediate regime, which interpolates regimes 1) and 3).

Our asymptotical function f(x) can be mapped into their three regimes' domain or vice versa. The first will correspond to our e+, e-, u, d –quarks, Walcher's (23 MeV), Mac Gregor-Akers-Palazzi (35 MeV, 70 MeV) goldstones, and a pion (135 MeV). The second regime will be home for our new X (214 MeV) particle. Third regime will be started from the 2.4 GeV = $\Lambda_{rest}$ and will go all the way to E = ∞. Many particles and resonances can live in II and III domains.

Let's take a closer look at light unflavored mesonic sector (the biggest of all). We can foresee such possible parity doublets (PD) there:

$$\begin{cases} \rho(1^+(J^{--})) \ - \ a(1^-(J^{++})) \ - \text{they are vector-axial vector doublets (VAV)} \\ h(0^-(J^{+-})) \ - \ \eta(0^+(J^{-+})) \\ f(0^+(J^{++})) \ - \ \omega(0^-(J^{--})) \\ f(0^+(J^{++})) \ - \ \varphi(0^-(J^{--})) \\ b(1^-(J^{+-})) \ - \ \pi(1^-(J^{-+})) \end{cases} \quad (3)$$

We can also create P-doublets via different symmetry principles. Now we have found the possible P-doublets in the light unflavored mesonic sector. These are f-ω and ρ-a -families. Let's consider the overlap value for different pairs in the quartets:

$$\begin{cases} f_1(1518) \ - \ \omega_1(1649) \\ a_1(1640) \ - \ \rho_1(1723) \end{cases} \quad - \ 12\% \text{ overlap}$$

$$\begin{cases} \omega_3(1955) \ - \ f_3(2048) \\ \rho_3(1994) \ - \ a_3(2070) \end{cases} \quad - \ 71\% \text{ overlap} \quad (4)$$

$$\begin{cases} \omega_3(2250) \ - \ f_3(2303) \\ \rho_3(2186) \ - \ a_3(2310) \end{cases} \quad - \ 100\% \text{ overlap}$$

The first quartet has $M_{ave1} = 1633$ MeV, the second quartet has $M_{ave2} = 2017$ MeV, and the third quartet has $M_{ave3} = 2262$ MeV. We are witnessing here an outstanding phenomenon, as we move higher in the spectrum. An overlap is getting larger and reaches 100% for the highest two pairs. What is more is that high in the spectra, ALL parity doublets become degenerate and fused into one state. This is exactly a harbinger for chiral symmetry restoration high in the mass spectrum! We note in passing that the occurrence of our 1, 2, 3 quartets resembles nicely the leveling-off behavior of our fitting function f(x).

Now we can combine these multiplets with so-called $U(1)_A$ pairs, originating from axial symmetry,

$$\begin{cases} \eta(2190) - f_0(2197) \\ \eta(2303) - f_0(2329) \\ \eta_2(2258) - f_2(2231) \\ \pi_4(2250) - a_4(2280) \end{cases} \quad (5)$$

From this set (5) one can see the creation of f-a-π-η box with boundaries 2190-2329 MeV. Let us combine set (5) with vector-axial-vector ρ-a parity doublets, which will confine within a 2186-2329 MeV box. As a result, we have a high-lying supermultiplet, exhibiting a perfect match between VAV parity doublets and $U(1)_A$ pairs, with box 2186-2329 MeV. This proves total degeneracy of states and restoration of the chiral symmetry high in the spectrum.

On the other hand our extensive research on hadronic Regge trajectories (RT) shows [16] that our asymptotic fitting curve f(x) could be well matched by the bulk of the mesonic and baryonic RT's in orbital space, i.e. $M^2 = M^2(L, n_r = \text{fixed})$. Our asymptote boundary at $M = 2.4$ GeV beautifully corresponds to the flattening out of RT's. We know that clearly light unflavored mesons and N - Δ - Λ - Σ baryonic spectra do terminate at 2.4 – 2.5 GeV value (and most definitely by 3 GeV) [17].

## V. Conclusions

To this date, no one has been able to successfully explain the mass spectrum of all the known particles. We don't pretend that we can fully explain it either but we expect the solution to be a geometrical one of some fashion. Our 2D Spin Matrix geometry is our first attempt for a geometrical solution and it has some promising features that warrant further investigation. Starting with the rest mass energy of the electron or positron, our geometry generates energy states that can be applied to the up and down quarks, pi zero mesons, the new HyperCP particle and many others. We have to stress that our scheme is parameter-free.

We can also see areas of chiral symmetry as well as areas of other symmetrical principles. A curious symmetry can be seen for the LTV $d\bar{d}$ quark areas (8.111 MeV) in one of the isolated regions. It seems that these three $d\bar{d}$ quark areas would have to have a three-way chirality with respect to each other. Normally one thinks of left-handed and right-handed for chiral symmetries, but we are wondering what would happen if a three-way chirality was applied to the group theories of particle physics. Another curious symmetry is that either the LTV $e^+e^-$ states or the LTV $u\bar{u}$ quark states can be the center of symmetry for our Spin Matrix geometry.

In presenting this geometrical model, it is easy to see why it has been so hard to find an explanation and mathematical formulation for the particle spectrum. The particle spectrum is dependent on an equilibrium of complicated field interactions in the *less than virtual* sector of the quantum vacuum. Perhaps our Spin Matrix is a start to finding the proper explanation.